\documentclass{epl}
\usepackage{amsmath}
\usepackage{amssymb}
\title{Density functional theory of freezing for soft interactions in two dimensions}
\shorttitle{Density functional theory of freezing in 2D}
\author{S.~van Teeffelen\thanks{E-mail: \email{teeffelen@thphy.uni-duesseldorf.de}} \and C.~N.~Likos \and N.~Hoffmann \and H.~L\"owen}
\shortauthor{S.~van Teeffelen \etal}
\institute { Institut f\"ur Theoretische Physik II,
Heinrich-Heine-Universit\"at D\"usseldorf,\\ Universit\"atsstra{\ss}e 1,
D-40225 D\"usseldorf, Germany}


\begin{document}

\maketitle

\begin{abstract}
  A density functional theory of two-dimensional freezing is presented
  for a soft interaction potential that scales as inverse cube of
  particle distance.  This repulsive potential between parallel,
  induced dipoles is realized for paramagnetic colloids on an
  interface, which are additionally exposed to an external magnetic
  field.  An extended modified weighted density approximation which
  includes correct triplet correlations in the liquid state is used.
  The theoretical prediction of the freezing transition is in good
  agreement with experimental and simulation data.
\end{abstract}
%
%
A microscopic theory of freezing and melting is a great challenge in
statistical physics.  There are two complementary approaches to the
liquid-to-solid transition: first, classical density functional
theory~\cite{Oxtoby:90,Loewen:94,Singh:91} starts from liquid state
and views the solid as a condensation of liquid density modes, hence
it is a liquid-based approach.  Second, crystal elasticity
theory~\cite{Strandburg:88} is a solid-based theory where the liquid
is viewed as a solid with an accumulation of defects.  In three
dimensions, the freezing transition is first order and it is known
that it is not defect mediated. Here, density functional theory
provides a molecular theory for the freezing transition. Crystal
elasticity theory is appropriate to two dimensions and predicts a
possible scenario of two-stage melting via an intermediate hexatic
phase \cite{Kosterlitz:73,Halperin:78,Nelson:79,Young:79}. The
advantage of density functional theory is that it can be used to
calculate the structure of the solid, whereas it is not possible to
extract the structure of the fluid out of crystal elasticity theory.

An excellent realization of a two-dimensional system is provided by
paramagnetic colloidal particles in a pendant water droplet, which are
confined to the air-water interface~\cite{Zahn:97}.  If an external
magnetic field is applied perpendicular to the interface, a magnetic
moment is induced in the particles resulting into a tunable, mutual
dipolar repulsion between them.  The corresponding interaction pair
potential $u(r)$ is repulsive and soft, being proportional to $1/r^3$,
with $r$ denoting the distance between the particles. The prefactor
can easily be tuned by varying the external magnetic field strength.
In real-space experiments~\cite{Zahn:99,vonGruenberg:04}, the
two-stage melting process was confirmed with an intermediate hexatic
phase which had a tiny stability range bracketed between the fluid and
crystalline phase.  There are also computer
simulations~\cite{Haghgooie:05,Loewen:96} for freezing in $1/r^3$
system but finite-size effects turn out to become crucial
here~\cite{Binder:02:Bates:00}.

In this letter, we apply density functional theory (DFT) to
two-dimensional freezing of soft $1/r^3$ interactions. There are two
major difficulties arising in doing so: first, it is known that it is
difficult to get a reliable density functional approximation for soft
repulsive interaction potentials. While hard sphere freezing serves as
standard test case for various approximations and many reliable
approximations do exist (e.g. Rosenfeld's fundamental measure
theory~\cite{Rosenfeld:89}), the freezing of soft inverse-power
law-fluids turns out to be much harder~\cite{wang:gast:jcp:99}. For
the extreme case of the one-component plasma, featuring a $1/r$
interaction, it has been shown by Likos and
Ashcroft~\cite{Likos:92:Likos:93} that an extended modified weighted
density approximation (EMA) which contains correct triplet correlation
of the fluid yields reliable freezing data. In this letter, we
overcome the first problem in a similar way and generalize the EMA to
two dimensions and apply it to the $1/r^3$ interaction. The second,
more principal problem is linked to the two-dimensional character of
the system itself. It is not clear to date how to include the hexatic
phase into the density functional language. Here, we do not address
this deep question but rather focus on the prediction of the freezing
transition point neglecting the tiny stability range of the hexatic
phase. A similar view has been taken for hard disk
systems~\cite{Zeng:90,Barrat:88b,Xu:90,Ryzhov:95} and to the $1/r$
interaction in 2D~\cite{Ryzhov:95} where density functional theory of
freezing was applied to.

We demonstrate that the EMA yields excellent freezing data as compared
to the standard modified weighted density approximation
(MWDA)~\cite{Denton:89b}. Even the relative-mean-square-displacement
in the coexisting solid is in reasonable agreement with the
experimental data on colloids in a magnetic field.

The Helmholtz free energy density functional is typically split into
the ideal gas and an excess part, $ F_{\rm{tot}}\left[\rho({\bm
    r})\right]= F_{\rm{id}}\left[\rho({\bm r})\right]
+F_{\rm{ex}}\left[\rho({\bm r})\right]$. Here the ideal part is local
and nonlinear, $F_{\rm{id}}\left[\rho({\bm r})\right]= k_B T
\int\,\upd {\bm r} \rho({\bm r})\left\{ \ln\left[\rho({\bm
      r})\Lambda^3\right]-1\right\}$, with $\Lambda$ denoting the
thermal de Broglie wavelength and $k_B T$ the thermal energy.  The
excess part can only be calculated approximately.  Both the MWDA and
the EMA approximate the excess free energy of the inhomogeneous system
by setting it equal to the excess free energy of a uniform liquid
evaluated at an appropriately weighted density $\hat{\rho}$:
\begin{equation}\label{eq:fmwda}
F_{\rm{ex}}\left[\rho({\bm r})\right]\approx
F_{\rm{ex}}^{\rm{MWDA/EMA}}\left[\rho({\bm r})\right]
=N f_0(\hat\rho),
\end{equation}
where $N$ is the number of particles in the system and $f_0(\hat\rho)$
is the excess free energy per particle of the uniform liquid at the
weighted density
\begin{align}\begin{split}\label{eq:rhohat}\hat{\rho}\left[\rho ({\bm r})\right]=
  &  \frac{1}{N}\int\,\upd {\bm r} \, \upd
    {\bm r}^\prime \rho({\bm r})\rho({\bm r}^\prime) w\left({\bm r}-{\bm
      r}^\prime;\hat{\rho}\right) \\
&+  \frac{1}{N^2}\int\,\upd {\bm r} \, \upd {\bm r}^\prime \, \upd {\bm r}^{\prime\prime}
  \rho({\bm  r})\rho({\bm r}^\prime)\rho({\bm r}^{\prime\prime}) v\left({\bm
      r}-{\bm r}^\prime,{\bm r}-{\bm
      r}^{\prime\prime};\hat{\rho}\right).
\end{split}\end{align}
Here the second term only appears in the EMA and not in the MWDA.  The
weight functions $w({\bm r};\rho)$ and $v({\bm r},{\bm
  r}^\prime;\rho)$ are determined by requiring that in the uniform
limit, $\rho({\bm r})\rightarrow\rho $ the approximate functional
$F_{\rm{ex}}\left[\rho({\bm r})\right]$ is exact up to second (MWDA)
or third (EMA) order in density difference $\Delta\rho({\bm
  r})=\rho({\bm r})-\rho$ in the functional expansion of the excess
free energy of the inhomogeneous system about the excess free energy
of the fluid.  The second weight function $v$ in eq.~(\ref{eq:rhohat})
is chosen to be zero in the MWDA since the third order correlation
function does not enter the formalism explicitly; rather, all higher
terms are approximately included as a consequence of the
self-consistency requirement on the determination of $\hat\rho$,
appearing on both sides of eq.~(\ref{eq:rhohat}).  The EMA, on the
other hand, is exact up to third order and, similarly, includes
approximate contributions from all higher-order terms.  The normalized
weight functions have to fulfill the
requirements~\cite{Likos:92:Likos:93}
\begin{align}\begin{split}\label{eq:deltafexnachdeltarho}
\lim_{\rho({\bm r})\rightarrow\rho } \left[\frac{\delta^2 F_{\rm{ex}}^{\rm{MWDA/EMA}}}{\delta \rho({\bm r})\delta \rho({\bm
  r}^\prime)}\right]&=-k_B T c_0^{(2)}\left({\bm r} - {\bm r}^\prime; \rho \right)\\
\lim_{\rho({\bm r})\rightarrow\rho } \left[\frac{\delta^3 F_{\rm{ex}}^{\rm{EMA}}}{\delta \rho({\bm r})\delta \rho({\bm
  r}^\prime)\delta \rho({\bm
  r}^{\prime\prime})}\right]&=-k_B T c_0^{(3)}\left({\bm r} - {\bm
r}^\prime, {\bm r} - {\bm r}^{\prime\prime};\rho \right)\,,
\end{split}\end{align}
where $c_0^{(2)}$ and $c_0^{(3)}$ are the two- and three-particle
direct correlation functions of the liquid~\cite{hansen-mcdonald:86}
which are an input to the theory.  These conditions uniquely determine
the weight functions and lead to simple algebraic equations for $v$
and $w$ that can be found in reference~\cite{Likos:92:Likos:93}.

In order to find the equilibrium one-particle density
$\rho_{\rm{eq}}({\bm r})$ we minimize the total free energy functional
$F_{\rm{tot}}^{\rm{MWDA/EMA}}\left[\rho({\bm r})\right]$ with respect to
the inhomogeneous one-particle density $\rho({\bm r})$. We make the
Gaussian ansatz, $\rho({\bm r})= \frac{\alpha}{\pi}\sum_{{\bm R}_i}
\exp\left[-\alpha\left|{\bm r} - {\bm R}_i \right|^2\right]$ where $\{{\bm
R}_i\}$ is the set of Bravais lattice vectors of a triangular lattice
(with average density $\rho$). For fixed average density $\rho$ we
thus end up with only one minimization parameter $\alpha$ which
describes the degree of localization.  For $\alpha\rightarrow 0 $ the
density profile becomes flat and the system turns into a homogeneous
liquid of number density $\rho$, whereas increasing $\alpha$ leads to
enhanced particle localization around the lattice sites.

With the Gaussian parametrization of the density profiles, we obtain
the following self-consistent equation for the weighted density
$\hat\rho$ as function of the localization parameter $\alpha$ and the
average density $\rho$:
\begin{equation}\label{eq:rhohatofrhos}
\frac{\hat\rho(\rho,\alpha)}{\rho}
=\left[1-\frac{k_BT}{2 f_0^\prime(\hat\rho)}\sum_{{\bm K}\neq
  0}\mu_K^2 \tilde c_0^{(2)}({\bm K};\hat\rho)
-\frac{\rho k_BT}{6 f_0^\prime(\hat\rho)}
\sum_{\substack{{\bm K}\neq {\bm 0}\\ {\bm Q}\neq {\bm 0},-{\bm K}}}
\mu_K \mu_Q \mu_{\left|{\bm K}+{\bm Q}\right|}
\tilde c_0^{(3)}({\bm K},{\bm Q};\hat\rho)\right]\,,
\end{equation}
where $\mu_k=e^{-k^2/4\alpha}$ are the Fourier coefficients of the
Gaussian ansatz for $\rho({\bm r})$ and, likewise, $\tilde c_0^{(n)}$,
$n = 2,3$, denote the Fourier transforms of the $n$-particle direct
correlation functions, evaluated at the reciprocal lattice vectors
(RLV's) ${\bm K}$ and ${\bm Q}$.  Primes denote derivatives with
respect to density and the three-particle term only appears in the
EMA.  Whereas $F_{\rm id}$ grows with $\alpha$, $\hat\rho$ decreases
with the latter, causing a concomitant decrease in $F_{\rm ex}$,
because the latter is a monotonically increasing function of
$\hat\rho$. As can be induced from eq.\ (\ref{eq:rhohatofrhos}) above,
the decrease of $\hat\rho$ with $\alpha$ is pronounced when the RLV's
of the lattice lie close to the maxima of $\tilde
c_0^{(2)}(k;\hat\rho)$, a feature that corresponds physically to an
inherent tendency of the fluid to enhance density waves at these
wavevectors.

We now apply the MWDA/EMA to the inverse-power pair potential
$u(r)=u_0/r^3$, where $u_0$ is a parameter with dimensions of energy
$\times$ volume; for the specific realization of two-dimensional
paramagnetic colloids of susceptibility $\chi$ exposed to a
perperndicular magnetic field ${\bf B}$, we have $u_0 = (\chi {\bf
  B})^2/2$. The thermodynamics and structure depend, due to simple
scaling, only on one relevant dimensionless coupling parameter $\Gamma
= u_0 \rho^{3/2}/(k_BT)$. Therefore, it is convenient to express all
quantities in terms of $\Gamma$ and consider coupling parameters
rather than densities via this scaling relation.  Correspondingly, in
what follows we employ the weighted coupling constant $\hat\Gamma$,
related to $\hat\rho$ via the scaling relation
$\hat\Gamma(\Gamma,\alpha)= u_0 \hat \rho^{3/2}(\rho,\alpha)/(k_{\rm
  B}T)$.

\begin{figure}
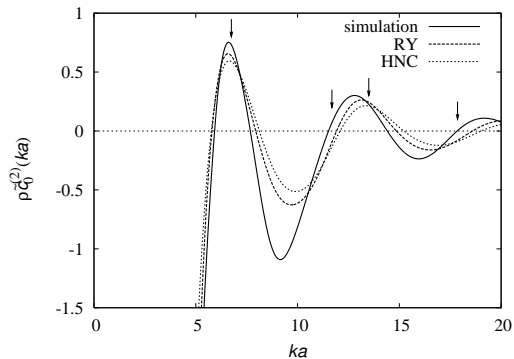

  \onefigure[width=7cm]{c2.epsf}
  \caption{The Fourier transform $\rho \tilde c_0^{(2)}(k)$
    of the two-particle direct correlation
    function at $\Gamma = 9$,
    plotted against $ka$, where $a = \rho^{-1/2}$. Shown are
    simulation data using the Verlet closure
    (solid line); liquid integral equation theory using the RY
    closure (dashed line) and liquid integral equation theory using
    the HNC closure (dotted line). The arrows indicate the positions of
    the first four reciprocal lattice vectors of the triangular
    lattice.}
  \label{fig:c2_ry_bd_hnc_g9}
\end{figure}

In order to obtain the concrete form of the functional approximations,
we need the two- and three-particle direct correlation functions and
the excess free energy per particle $f_0$ of the corresponding uniform
fluid for a wide range of coupling constants $\Gamma$. These
quantities are obtained as described below:

(i) The two-particle direct correlation function is obtained by liquid
integral equation theory, where we used the hypernetted chain
(HNC)~\cite{hansen-mcdonald:86} or the thermodynamically consistent
Rogers-Young (RY) closure~\cite{ry:pre:84}.  We have also produced
``exact'' data for the real-space direct two-particle correlation by
computer simulation using the Verlet closure~\cite{Verlet:68}.  A
comparison between the HNC, RY and simulation data for the Fourier
transform $\tilde{c}^{(2)}_{0}$ of the direct correlation function is
shown in fig.~\ref{fig:c2_ry_bd_hnc_g9} for the experimentally
determined coupling close to freezing.
The HNC closure underestimates the structure strongly while the RY
closure is closer to the simulation data. We also show the positions
of the first four reciprocal lattice vectors of a triangular lattice
with same density.  The value of $\tilde{c}^{(2)}_{0}$ at these
lattice vectors crucially influences the solid free energies, as can
be seen from eqs.~(\ref{eq:fmwda}) and (\ref{eq:rhohatofrhos}).
Therefore it can be anticipated that the HNC closure will not produce
reliable results and we do not consider it further.

(ii) The excess free energy per particle $f_0$ is obtained from the pair
correlation via the compressibility route.

(iii) Finally, the three-particle direct correlation function
$c_0^{(3)}$ of the underlying fluid is obtained from an approximation
by Denton and Ashcroft~\cite{Denton:89}, which is based on a weighted
density approximation to the first order direct correlation function
of an inhomogeneous system. This approach leads to an analytic
expression of $c_0^{(3)}$ in terms of the one- and two-particle
correlation functions $c_0^{(1)}$, $c_0^{(2)}$ of the liquid and their
derivatives with respect to density, employing the `symmetrized sum'
\begin{equation}
  \label{eq:c3da}
  \tilde{c}_0^{(3)}({\bm k}, {\bm k'})=\frac{1}{3}
\left[
 \tilde{f}\left(|{\bm k}|,|{\bm k}'|\right)
+\tilde{f}\left(|{\bm k}|,|{\bm k}+{\bm k'}|\right)
+\tilde{f}\left(|{\bm k}'|,|{\bm k}+{\bm k'}|\right)
\right],
\end{equation}
where
\begin{equation}
  \label{eq:c3da2}
  \tilde{f}(k,k')=\frac{1}{c_0^{(1)\prime}}
\left[\tilde c_0^{(2)}(k)\tilde c_0^{(2)\prime}(k')
+\tilde c_0^{(2)\prime}(k)\tilde c_0^{(2)}(k')\right]
-\frac{c_0^{(1)\prime\prime}}{\left[c_0^{(1)\prime}\right]^2}
\tilde c_0^{(2)}(k)\tilde c_0^{(2)}(k').
\end{equation}
Here, primes denote derivatives with respect to
density.

Results for the approximations proposed are presented in
figs.~\ref{fig:ghatofgs_emabd} and \ref{fig:f_emabd}.  In
fig.~\ref{fig:ghatofgs_emabd}, the weighted coupling constant
$\hat\Gamma(\Gamma,\alpha)$ is shown versus the localization parameter
$\alpha$ for a strong coupling $\Gamma$ close to freezing. Both the
MWDA and the EMA are examined with the simulation pair structure
input.  Obviously, $\hat\Gamma$ coincides with the bare $\Gamma$ in
the fluid ($\alpha=0$).  It can be seen that the MWDA yields a smaller
reduction in $\hat \Gamma$ relative to $\Gamma$ than the EMA: explicit
inclusion of three-body effects enhances the tendency of the particles
to localize. Hence, one expects freezing at lower couplings in the
EMA.  In fact, in fig.~\ref{fig:f_emabd}, the free energy difference
between a solid of localization $\alpha$ and a fluid ($\alpha=0$)
shown versus $\alpha$ reveals that the fluid is much more stable in
the MWDA as compared to the EMA.  The EMA yields a transition from the
fluid to the solid close to $\Gamma=9.4$: while for $\Gamma=9.0$ the
fluid is stable as indicated by the minimal value at $\alpha=0$,
fluid-solid coexistence is achieved at $\Gamma =9.4$, see the two
equal minima in fig.~\ref{fig:f_emabd}. The solid phase, on the other
hand, is clearly stable for $\Gamma=9.8$. The localization parameter
at coexistence is roughly $\alpha_{\rm min} a^2=100$.
\begin{figure}
  \twofigures[width=6.5cm]{ghatofalpha.epsf}{fofalpha.epsf}
  \caption{The weighted coupling constant $\hat \Gamma (\Gamma,
    \alpha)$ as a function of the localization parameter $\alpha$
    within the MWDA (solid line) and within the EMA (dashed line)
    using the ``exact'' pair structure from simulation for the strong
    coupling $\Gamma=9$.}
  \label{fig:ghatofgs_emabd}
  \caption{Relative free energy per particle $N^{-1}\left[F_{\rm{tot}}
      (\Gamma, \alpha)-F_{\rm{tot}}(\Gamma, \alpha=0)\right]$ in units
    of $k_B T$ as a function of the localization parameter $\alpha$
    obtained within the EMA for three different coupling constants
    $\Gamma=9,9.4,9.8$ (the three lower curves from top to bottom)
    compared to the relative free energy density obtained within the
    pure MWDA for a coupling constant $\Gamma=9$ (uppermost line).}
  \label{fig:f_emabd}
\end{figure}

The full results of a numerical calculation using Maxwell's double
tangent construction yield a weak first-order transition with a fluid
density corresponding to a coupling constant of $\Gamma_f$ and a solid
density corresponding to a coupling constant of $\Gamma_s$.  There is
a small density gap $\Delta \Gamma = \Gamma_s -\Gamma_f$, describing
the coexistence region. Table I summarizes the freezing/melting
parameters for the MWDA with RY closure, for the EMA with RY closure,
and for the EMA with the ``exact'' pair structure obtained from
simulation. The data are compared against experimental results
obtained from real-space microscopy measurement of magnetic colloids
confined to an air-water interface. The experiments give freezing with
an intermediate hexatic phase. The liquid-solid transition has also
been studied using numerical simulation~\cite{Haghgooie:05,Loewen:96}
yielding a slightly higher inverse transition temperature between
$12.0$ and $12.25$ but these investigations suffer from finite size
effects.

As becomes evident from Table I, the MWDA is not a quantitatively
satisfying theory as it overestimates the freezing coupling by a
factor of 4.  Note that the overestimation of the freezing coupling is
the reason why it is not possible to feed the ``exact'' pair structure
into the MWDA. At such high coupling, no fluid pair structures are
available since the fluid spontaneously crystallizes in the
simulation.  The EMA, on the other hand, yields results in close
agreement with experimental data.

More detailed, structural information can be extracted from the
localization parameter of the coexisting solid.  For all
approximations used we find localization parameters at freezing in the
range $ 99 < \alpha_{\rm{min}}(\Gamma_f)a^2 < 115$.  Strictly
speaking, the localization parameter has no counterpart in ``real'' 2D
systems since the particles are not localized due to long range
fluctuations.  However, if one relates the particle displacements to
that of their nearest neighbor, one can define a finite quantity as
$\gamma=\rho\left\langle \left({\bm u}_i - {\bm
      u}_{i+1}\right)^2\right\rangle$, where ${\bm u}_i$ and ${\bm
  u}_{i+1}$ are the displacement vectors of neighboring lattice sites.
Disregarding nearest-neighbor correlations $\left\langle {\bm u}_i
  \cdot {\bm u}_{i+1}\right\rangle$, $\gamma$ can be estimated. Since
the nearest-neighbor correlations $\left\langle {\bm u}_i \cdot {\bm
    u}_{i+1}\right\rangle$ are expected to be positive:
\begin{equation}
  \label{eq:lindemannratio}
  \gamma\lesssim 2 \rho \left\langle {\bm u}_i^2\right\rangle
  \approx 2/(\alpha_{\rm{min}}a^2).
\end{equation}
By this relation, the localization parameter of the coexisting solid
gives a prediction for $\gamma$ which is included in Table I.  From
experiments, $\gamma$ is known to be close to $\cong
0.038$~\cite{Zahn:99}. This was shown to be in accordance with
harmonic lattice theory~\cite{Froltsov:05}.  The EMA yields
$\gamma\lesssim 0.020$, i.e.  the EMA roughly overestimates the
localization of the particles by a factor of $2$. $\gamma$ is {\it
  smaller} than the experimental value, contrarily to what was
expected from the inequality~(\ref{eq:lindemannratio}).  This shows
that there is still a need to improve the theories in order to
correctly predict localization properties. A similar overestimation of
the localization is also common in weighted density approximations in
three spatial dimensions~\cite{Denton:89b}.

\begin{table}
  \caption{Freezing and melting parameters $\Gamma_f$ and $\Gamma_s$, the 
    widths of the coexistence regions $\Delta\Gamma=\Gamma_s-\Gamma_f$, and the 
    relative displacement parameters $\gamma$ at coexistence obtained 
    within: the MWDA with the RY closure (first row); the EMA with the RY 
    closure (second row); the EMA with the ``exact'' 
    pair structure from simulation (third row) and experimental parameters for the 
    isotropic-hexatic transition, the hexatic-crystal transition, and the 
    Lindemann parameter, obtained from real-space microscopy measurements 
    of magnetic colloids confined to an air-water interface (last row).}
\label{tab:freezing}
\begin{center}\begin{tabular}{l|lllll}
    & $\Gamma_f$ & $\Gamma_s$ & $\Delta\Gamma$& $\gamma$\\
    \hline
    MWDA  with RY               & 41.07   & 41.13  & 0.06   & 0.017 \\
    EMA   with RY               & 23.0    & 23.08  & 0.09   & 0.020 \\
    EMA   with simulation       & 9.33    & 9.49   & 0.16   & 0.020 \\
    Experiment                  &    10.0  & 10.75  & -     & 0.038 \\
\end{tabular}\end{center}\end{table}

In conclusion, we have demonstrated that the EMA is able to
quantitatively predict the freezing transition of a two-dimensional
colloidal system with soft and long-ranged $1/r^3$-interactions in
good agreement with experimental and simulation data. In analogy to
three-dimensional systems, the appearance of long-range interactions
requires the explicit inclusion of three-particle correlation
functions of the liquid in the construction of the weighted density.
Furthermore, the predicted transition temperatures are very sensitive
towards slight changes of the two-particle correlation functions of
the underlying fluid. A highly accurate input of the same is therefore
crucial.

Relying on the good quality of the EMA functional, our results can
serve as a platform to treat more challenging problems than bulk
transitions~\cite{Loewen:01}.  One obvious extension is towards
external potentials acting on the particles, such as system walls or
gravity.  The MWDA can in principle be applied to a fluid near a
single wall, but not to a free interface between coexisting phases.
Another interesting example is a spatially inhomogeneous magnetic
field, which renders the interactions space
dependent~\cite{Froltsov:04}. Finally, one may employ the scheme of
dynamical density functional theory~\cite{Marconi:99,Dzubiella:03} in
order to use the EMA functional to study the effect of spatially
homogeneous magnetic fields that oscillate in time. Work along these
lines is currently under way.

\acknowledgments This work has been supported by the DFG within the
SFB~TR6, project section C3.

\end{document}